\documentclass[aps,prl,preprint,tightenlines,superscriptaddress,showpacs,byrevtex]{revtex4}

\usepackage{epsf}
\usepackage{graphicx} 
\usepackage{dcolumn}  

\graphicspath{{ps}}

\def\cp{$CP$\/}
\def\cpv{$CPV$\/}
\def\mkpi{$m^{}_{K\pi}$}

\def\mevm{~MeV/$c^2$\/}

\def\meve{~MeV}

\def\gevp{~GeV/$c$\/}

\def\ra{\!\rightarrow\!}
\def\kbar{\overline{K}{}^{\,0}}
\def\kkbar{$K^0$-$\kbar$}
\def\dbar{\overline{D}{}^{\,0}}
\def\ddbar{$D^0$-$\dbar$}
\def\bbar{\overline{B}{}^{\,0}}
\def\bbbar{$B^0$-$\bbar$}
\def\dkpi{$D^0\ra K^-\pi^+$}

\def\dkpiw{$D^0\ra K^+\pi^-$}
\def\dstar{$D^{*\,+}\ra D^0\pi^+$}

\begin{document}

\vspace*{-3\baselineskip}
\resizebox{!}{3cm}{\includegraphics{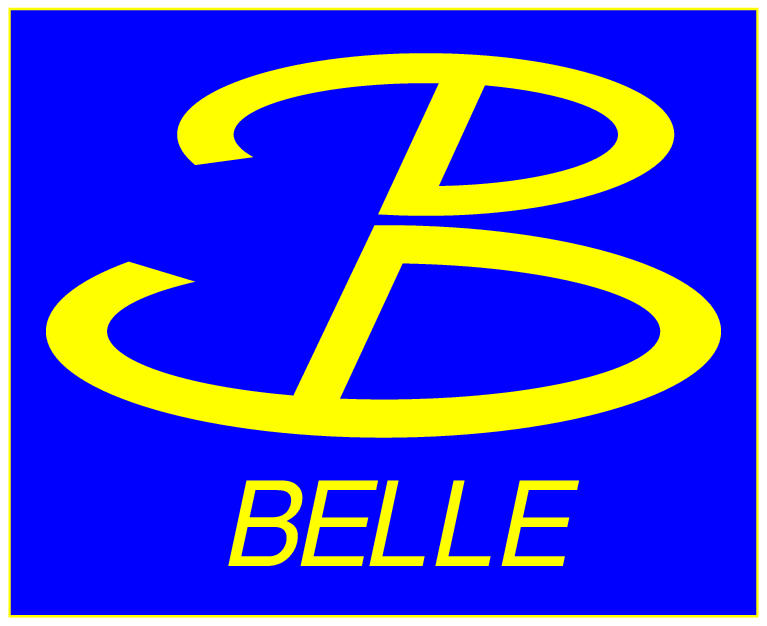}}

\preprint{\vbox{ \hbox{   }
                 \hbox{BELLE Preprint 2004-29}
                 \hbox{KEK Preprint 2004-58}
                 \hbox{UCHEP-04-02} 
}}

\title{ \quad\\[0.5cm]  
Search for \boldmath{$D^0$-$\dbar$} mixing in \boldmath{\dkpiw}
decays and measurement of the doubly-Cabibbo-suppressed decay
rate}

\affiliation{Budker Institute of Nuclear Physics, Novosibirsk}
\affiliation{Chiba University, Chiba}
\affiliation{Chonnam National University, Kwangju}
\affiliation{University of Cincinnati, Cincinnati, Ohio 45221}
\affiliation{University of Frankfurt, Frankfurt}
\affiliation{Gyeongsang National University, Chinju}
\affiliation{University of Hawaii, Honolulu, Hawaii 96822}
\affiliation{High Energy Accelerator Research Organization (KEK), Tsukuba}
\affiliation{Hiroshima Institute of Technology, Hiroshima}
\affiliation{Institute of High Energy Physics, Chinese Academy of Sciences, Beijing}
\affiliation{Institute of High Energy Physics, Vienna}
\affiliation{Institute for Theoretical and Experimental Physics, Moscow}
\affiliation{J. Stefan Institute, Ljubljana}
\affiliation{Korea University, Seoul}
\affiliation{Kyungpook National University, Taegu}
\affiliation{Swiss Federal Institute of Technology of Lausanne, EPFL, Lausanne}
\affiliation{University of Ljubljana, Ljubljana}
\affiliation{University of Maribor, Maribor}
\affiliation{University of Melbourne, Victoria}
\affiliation{Nagoya University, Nagoya}
\affiliation{Nara Women's University, Nara}
\affiliation{National Central University, Chung-li}
\affiliation{National United University, Miao Li}
\affiliation{Department of Physics, National Taiwan University, Taipei}
\affiliation{H. Niewodniczanski Institute of Nuclear Physics, Krakow}
\affiliation{Nihon Dental College, Niigata}
\affiliation{Niigata University, Niigata}
\affiliation{Osaka City University, Osaka}
\affiliation{Peking University, Beijing}
\affiliation{Princeton University, Princeton, New Jersey 08545}
\affiliation{University of Science and Technology of China, Hefei}
\affiliation{Seoul National University, Seoul}
\affiliation{Sungkyunkwan University, Suwon}
\affiliation{University of Sydney, Sydney NSW}
\affiliation{Tata Institute of Fundamental Research, Bombay}
\affiliation{Toho University, Funabashi}
\affiliation{Tohoku Gakuin University, Tagajo}
\affiliation{Tohoku University, Sendai}
\affiliation{Department of Physics, University of Tokyo, Tokyo}
\affiliation{Tokyo Institute of Technology, Tokyo}
\affiliation{Tokyo Metropolitan University, Tokyo}
\affiliation{Tokyo University of Agriculture and Technology, Tokyo}
\affiliation{University of Tsukuba, Tsukuba}
\affiliation{Virginia Polytechnic Institute and State University, Blacksburg, Virginia 24061}
\affiliation{Yonsei University, Seoul}
  \author{J.~Li}\affiliation{University of Science and Technology of China, Hefei} 
 \author{Z.~P.~Zhang}\affiliation{University of Science and Technology of China, Hefei} 
  \author{A.~J.~Schwartz}\affiliation{University of Cincinnati, Cincinnati, Ohio 45221} 
  \author{K.~Abe}\affiliation{High Energy Accelerator Research Organization (KEK), Tsukuba} 
  \author{K.~Abe}\affiliation{Tohoku Gakuin University, Tagajo} 
  \author{H.~Aihara}\affiliation{Department of Physics, University of Tokyo, Tokyo} 
  \author{M.~Akatsu}\affiliation{Nagoya University, Nagoya} 
  \author{Y.~Asano}\affiliation{University of Tsukuba, Tsukuba} 
  \author{T.~Aushev}\affiliation{Institute for Theoretical and Experimental Physics, Moscow} 
  \author{A.~M.~Bakich}\affiliation{University of Sydney, Sydney NSW} 
% ...
\collaboration{The Belle Collaboration}

\noaffiliation

\begin{abstract}
We have searched for mixing in the $D^0$-$\dbar$ system
by measuring the decay-time distribution of \dkpiw\ decays. 
The analysis uses 90~fb$^{-1}$ of data collected 
by the Belle detector at the KEKB $e^+e^-$ collider. We fit 
the decay-time distribution for the mixing parameters $x'$ 
and $y'$ and also for the parameter $R^{}_D$, which is the 
ratio of the rate for the doubly-Cabibbo-suppressed decay
\dkpiw\ to that for the Cabibbo-favored decay \dkpi. We 
do these fits both assuming \cp\ conservation and allowing 
for \cp\ violation. We use a frequentist method to obtain 
a 95\% C.L.\ region in the $x'^2$-$y'$ plane.
Assuming no mixing, we measure 
$R^{}_D=(0.381\,\pm\,0.017\,^{+0.008}_{-0.016})\%$.
\end{abstract}


\maketitle


{\renewcommand{\thefootnote}{\fnsymbol{footnote}}}
\setcounter{footnote}{0}

The phenomenon of
mixing among quark flavors has been observed
in the \kkbar\ and \bbbar\ systems but not yet 
in the \ddbar\ system. The rate for \ddbar\ mixing 
within the Standard Model (SM) is small, typically well 
below experimental upper limits~\cite{cicerone}. Observation 
of mixing much larger than this expectation could 
indicate new physics, such as a $\Delta C\!=\!2$ interaction.
Such nonstandard processes may also give rise to \cp-violating effects. 

In this paper we present a search for \ddbar\ mixing and 
\cp\ violation (\cpv) in mixing with greater sensitivity 
than that of previous searches. The data sample consists 
of 90~fb$^{-1}$ recorded by the Belle experiment 
at the KEKB asymmetric $e^+e^-$ collider~\cite{kekb}. 
The Belle detector~\cite{belle_detector} consists of 
a three-layer silicon vertex detector (SVD), a 50-layer
central drift chamber (CDC), an array of aerogel threshold 
Cherenkov counters (ACC), time-of-flight scintillation 
counters (TOF), and an electromagnetic calorimeter
comprised of CsI(Tl) crystals. These detectors are
located within a solenoid coil providing a 1.5~T 
magnetic field. An iron flux-return 
outside the coil is instrumented to identify 
muons and $K^0_L$ mesons.

The dominant two-body decay of the $D^0$ is the
Cabibbo-favored (CF) decay \dkpi~\cite{charge-conjugate}.
We search for mixing by reconstructing the ``wrong-sign'' decay 
\dkpiw, which would arise from a $D^0$ mixing to $\dbar$
and subsequently decaying via $\dbar\ra K^+\pi^-$.
The flavor of the $D$ is identified by requiring
that it originate from $D^{*\,+}\ra D^0\pi^+$ or
$D^{*\,-}\ra\dbar\pi^-$ and noting the charge of 
the accompanying pion. In addition to arising 
via mixing, \dkpiw\ can also occur via a doubly-Cabibbo-suppressed (DCS)
amplitude. The two processes can be distinguished via
the decay-time distribution. This method has 
been used by FNAL E791~\cite{e791}, CLEO~\cite{cleo}, 
and BaBar~\cite{babar} to search for
mixing and measure or constrain the DCS decay rate.

The parameters used to characterize mixing are
$x\equiv \Delta m/\overline{\Gamma}$ and
$y\equiv \Delta\Gamma/(2\overline{\Gamma})$, where
$\Delta m$ and $\Delta \Gamma$ are the differences in mass
and decay width between the two \ddbar\ mass eigenstates, 
and $\overline{\Gamma}$ is the mean decay width.
For $|x|,\,|y|\ll 1$ and negligible \cpv, the decay
time distribution for \dkpiw\ can be expressed as
\begin{eqnarray}
\hskip-0.20in \frac{dN}{dt} & \!\propto\! & 
e^{-\overline{\Gamma}\,t}
\left[ R^{}_D + \sqrt{R^{}_D}\,y' (\overline{\Gamma}t) + 
\frac{x'^2 + y'^2}{4} (\overline{\Gamma}t)^2 \right],
\label{eqn:time_dep}
\end{eqnarray}
where $R^{}_D$ is the ratio of DCS to CF decay rates,
$x' = x\cos\delta + y\sin\delta$, 
$y' = y\cos\delta - x\sin\delta$, and 
$\delta$ is the strong phase difference between 
the DCS and CF amplitudes.
The first term in brackets is due to the
DCS amplitude, the last term is due to 
mixing, and the middle term is due to 
interference between the two processes.
The time-integrated rate for \dkpiw\ 
relative to 
that for \dkpi\ is
$R^{}_D + \sqrt{R^{}_D}\,y' + \left(x'^2 + y'^2\right)/2$ .

To allow for \cp\ violation, we follow Ref.~\cite{babar} and
apply Eq.~(\ref{eqn:time_dep}) to $D^0$ and $\dbar$ decays 
separately. This results in six observables: 
$\{ R^+_D,\,x'^{+\,2}\!,\,y'^+ \}$ for $D^0$ and 
$\{ R^-_D,\,x'^{-\,2}\!,\,y'^- \}$ for $\dbar$.
\cp\ violation is parametrized by the asymmetries
$A^{}_D = (R^+_D - R^-_D)/(R^+_D + R^-_D )$
and 
$A^{}_M = (R^+_M - R^-_M)/(R^+_M + R^-_M )$, 
where 
$R^\pm_M = \left(x'^{\pm\,2} + y'^{\pm\,2}\right)/2$. 
The asymmetry $A^{}_D$ characterizes \cpv\ in the DCS decay 
amplitude, and $A^{}_M$ characterizes \cpv\ in \ddbar\ mixing. 
The observables are related to $x'$ and $y'$ via
\begin{eqnarray}
x'^\pm & = & \left[\frac{1\pm A^{}_M}{1\mp A^{}_M}\right]^{\frac{1}{4}}
		\left(x'\cos\phi\pm y'\sin\phi\right) \label{eqn:cpv1} \\
 & & \nonumber  \\
y'^\pm & = & \left[\frac{1\pm A^{}_M}{1\mp A^{}_M}\right]^{\frac{1}{4}}
		\left(y'\cos\phi\mp x'\sin\phi\right)\,,
\label{eqn:cpv2}
\end{eqnarray}
where $\phi$ is a weak phase and characterizes \cpv\ occurring 
via interference between mixed and unmixed decay amplitudes. 
Note that $x'^\pm,\,y'^\pm$ are
unchanged by the transformation $x'\ra -x'$, $y'\ra -y'$,
and $\phi\ra\phi+\pi$; thus for definiteness we restrict
$\phi$ to the range $\left|\phi\right|<\pi/2$.

We select \dkpi\ decays by requiring two oppositely-charged 
tracks, with at least four SVD hits, 
that satisfy $K$ and $\pi$ identification criteria. These criteria
are ${\cal L}^{}_K\!>\!0.6$ and ${\cal L}^{}_\pi\!>\!0.4$, where 
${\cal L}$ is the relative likelihood for a track to be a $K$ or $\pi$ 
based on $dE/dx$ information in the CDC and the responses of the 
TOF and ACC systems. These criteria have efficiencies of 88.0\% 
and 88.5\% and $\pi/K$ misidentification rates of 8.5\% and 8.8\%,
respectively. We combine the $D^0$ candidate with a low-momentum pion 
($\pi^{}_{\rm slow}$) to form a \dstar\ candidate. Candidates in which 
the charge of $\pi^{}_{\rm slow}$ is opposite (equal to) that of the 
$K^\pm$ are referred to as ``right-sign'' or RS (``wrong-sign'' or WS) 
decays. To reject WS background from \dkpi\ in which the $K$ is 
misidentified as $\pi$ and the $\pi$ 
is misidentified as $K$, 
we recalculate $m^{}_{K\pi}$ with the $K$ and $\pi$ assignments 
swapped and reject events with 
$|\,m^{}_{K\pi}\!\!-m^{}_{D^0} |< 28$\mevm.
To eliminate $D^*$'s from $B$ decays, we require 
$p^{}_{D^*}>2.5$\gevp, where $p^{}_{D^*}$ is evaluated 
in the $e^+e^-$ center-of-mass frame.

The $D^0$ vertex is obtained by fitting the daughter $K/\pi$
tracks. The $D^*$ vertex is taken as the intersection of the 
$D^0$ momentum vector with the interaction profile 
region. We require a good $\chi^2$ for each vertex fit. 
The momentum of $\pi^{}_{\rm slow}$ is refitted with
the constraint that it originates from the $D^*$ vertex.
The $D^0$ decay time is calculated as 
$(\ell^{}_{D^0}/p^{}_{D^0})\times m^{}_{D^0}$, where 
$\ell^{}_{D^0}$ is the distance between the $D^0$ and $D^*$
vertices projected onto the $\vec{p}^{}_{D^0}$ direction.
The decay-time resolution is typically~0.2~ps.

We measure $R^{}_D,\,x'^2\!$, and $y'$ 
of Eq.~(\ref{eqn:time_dep}) via an unbinned maximum likelihood 
fit to the WS decay-time distribution. The likelihood function
consists of probability density functions (pdf's) for signal and
several backgrounds. The pdf's depend on the decay time, the 
mass $m^{}_{K\pi}$, and the kinetic energy released 
$Q\equiv m^{}_{K\pi\pi^{}_{\rm slow}}\!\!\!-m^{}_{K\pi}\!-m^{}_{\pi}$.
The latter equals only 5.85\mevm\ for 
$D^{*\,+}\ra D^0\pi^+_{\rm slow}\ra K\pi\pi^+_{\rm slow}$ decays,
which is near the threshold.

The signal pdf for event $i$ is smeared by a resolution function 
$R^{}_i = \left(1-f^{}_{\rm tail}\right)\,G(t^{}_i-t', 
\sigma^{}_{t,i}\ ; \mu, S) + 
f^{}_{\rm tail}\,G(t^{}_i-t', \sigma^{}_{t,i}\ ; \mu, S^{}_{\rm tail})$,
where the $G$'s are Gaussians with common mean $\mu$ and standard
deviations $(S\times\sigma^{}_{t,i})$ and
$(S^{}_{\rm tail}\times\sigma^{}_{t,i})$, and $\sigma^{}_{t,i}$ is
the uncertainty in decay time~$t$ for event~$i$. The parameters
$f^{}_{\rm tail},\,\mu$, and scaling factors $S$ and $S^{}_{\rm tail}$ 
are determined from data. The background pdf's are smeared by
similar resolution functions (see below).
To check the resolution function, we fit the RS sample 
in the same manner as the WS sample except that the 
signal pdf has a purely exponential time dependence.

\begin{figure}
\centerline{\epsfxsize=7.0cm \epsfbox{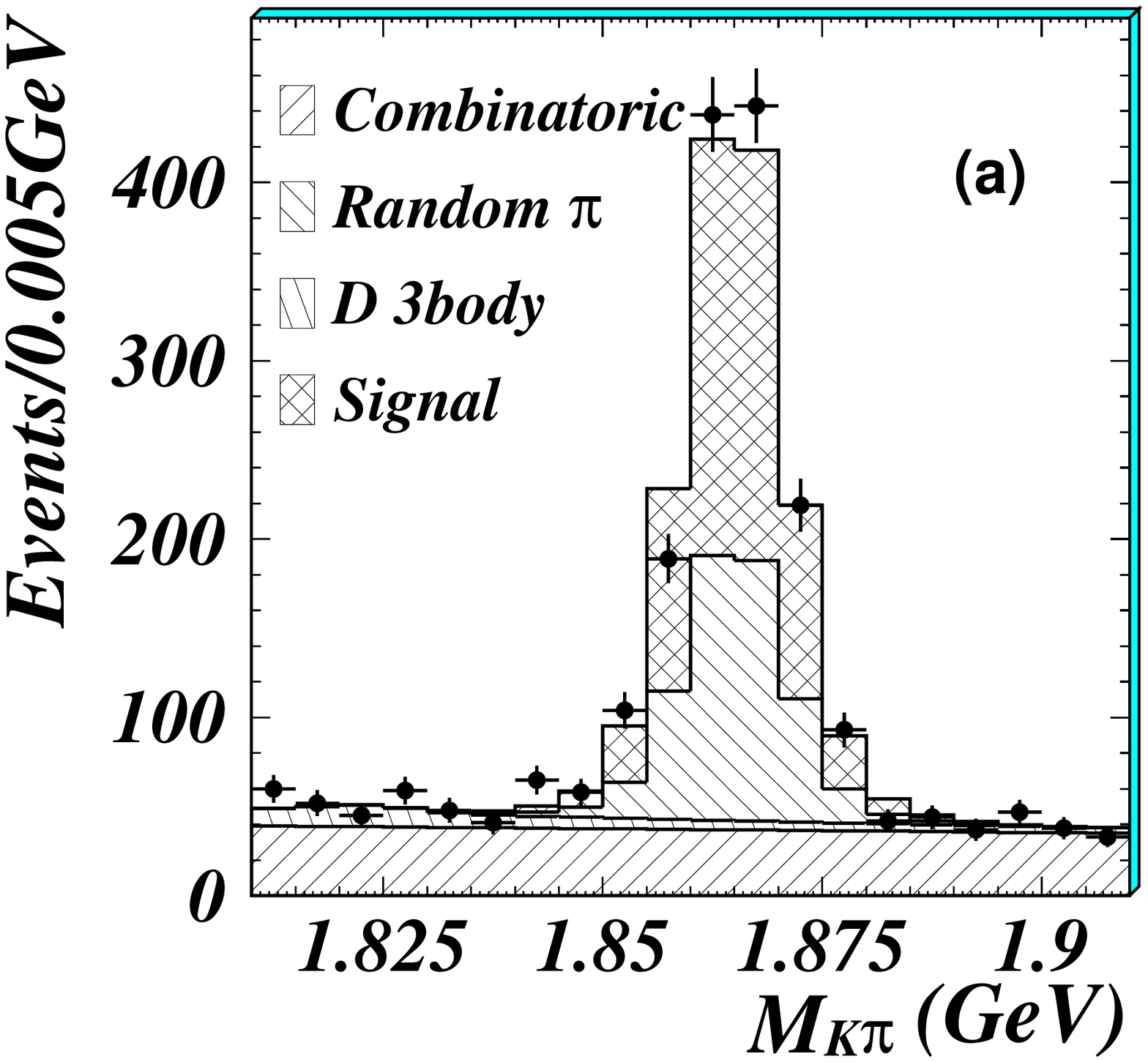}\hskip0.30in
            \epsfxsize=7.0cm \epsfbox{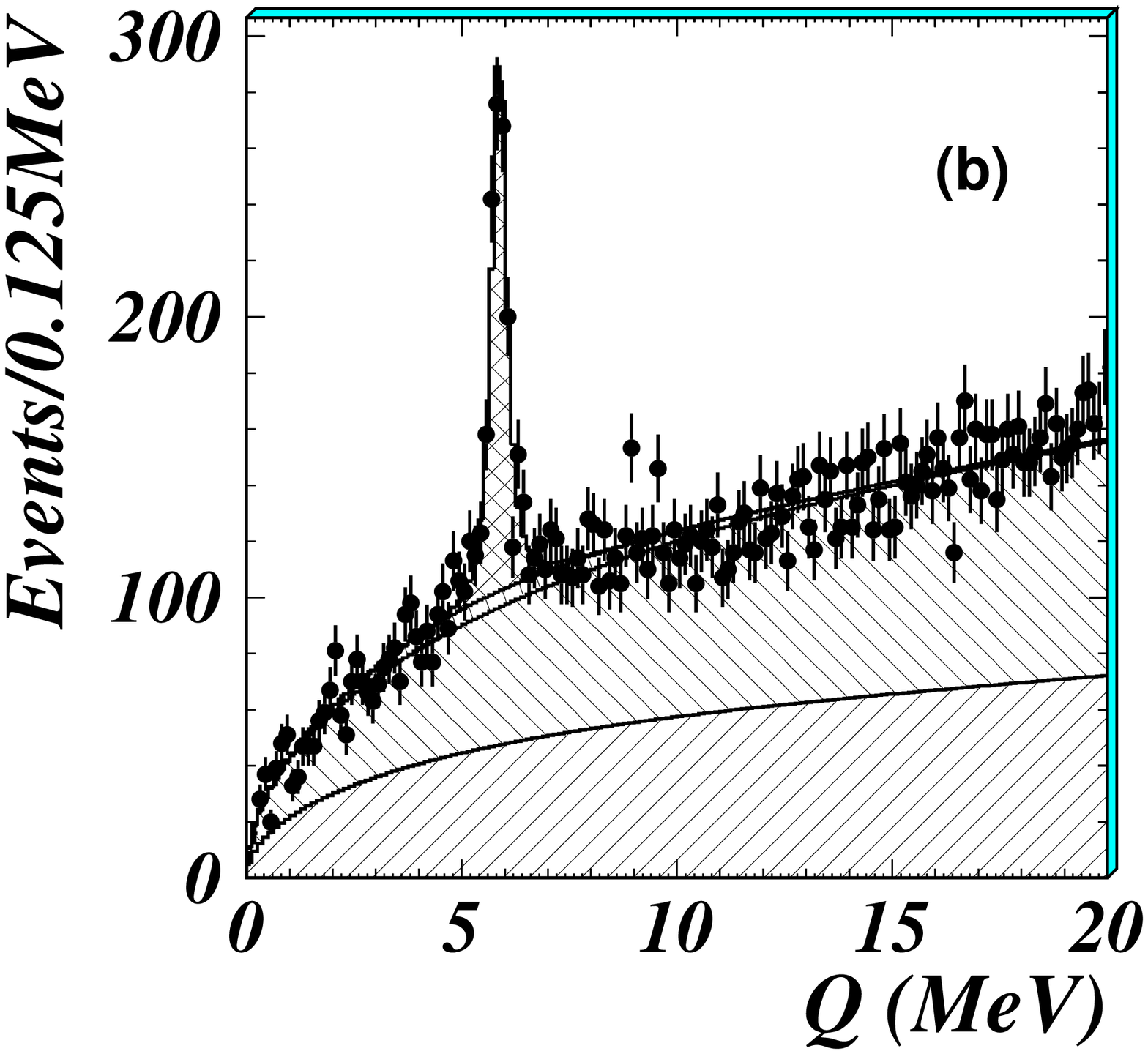}}
\vskip-0.05in
\caption{Distributions of
{\it (a)\/} WS \mkpi\ with $\left|\,Q-5.9{\rm\ MeV}\,\right|\!<\!0.6$\meve; and
{\it (b)\/} WS $Q$ with $\left|\,m^{}_{K\pi}-m^{}_{D^0}\right|\!<\!20$\mevm.
Superimposed on the data (points with error bars) are projections 
of the \mkpi-\,$Q$ fit.}
\label{fig:samples}
\end{figure}

There are four backgrounds to the WS sample:
{\it (a)\/} random $\pi$ background, in which a random $\pi^+$ 
is paired with a $\dbar\ra K^+\pi^-$ decay
(the pdf is peaked in \mkpi\ but broad in $Q$);
{\it (b)\/} \dstar\ followed by $D^0$ decaying to $\geq$\,3-body
final states (the pdf is broad in \mkpi\ and broad but
enhanced in $Q$);
{\it (c)\/} $D^+/D^+_s$ decays; and
{\it (d)\/} combinatorial.
We determine the level of each background 
by performing a two-dimensional fit to the \mkpi-\,$Q$ 
distribution. When fitting the RS sample, the \mkpi\ and $Q$ 
means and widths for signal are floated; when fitting the WS sample, 
these means and widths are fixed to the values obtained from
the RS fit. Also for the WS fit, the ratio of $D^+/D^+_s$ 
to $\geq$\,3-body backgrounds is fixed 
to the value obtained from Monte Carlo (MC) simulation. 
The RS fit finds $227\,721\pm 497$ \dkpi\ decays, 
and the WS fit finds $845\pm 40$ \dkpiw\ decays. 
The ratio $R^{}_{\rm WS}\equiv \Gamma(D^0\ra K^+\pi^-)/\Gamma(D^0\ra K^-\pi^+)
= (0.371\pm 0.018)$\% (statistical errors only).
The ratio of WS signal to background is~0.9; the latter
is mostly random~$\pi$ (59\%) and combinatorial~(36\%). 
The WS 
\mkpi\ and $Q$ distributions are shown in 
Fig.~\ref{fig:samples}
along with the fit projections.

To fit the decay-time distributions of RS and WS samples, 
we consider the $4\sigma$ region 
$\left|m^{}_{K\pi}-m^{}_{D^0}\right|\!<\!22$\mevm\ 
and $\left|\,Q - 5.9{\rm\ MeV}\right|\!<\!1.5$\mevm.
The signal and background yields (which normalize the
pdf's in the likelihood function) are 
determined from the \mkpi-\,$Q$ fit described above.
The decay-time distributions for the backgrounds before 
smearing are taken to be:
$e^{-t/\tau^{}_{D^0}}$ for random $\pi$ background,
$e^{-t/\tau^{}_{D3b}}$ for $\geq 3$-body $D^0$ background, 
$e^{-t/\tau^{}_{Dch}}$ for $D^+/D^+_s$ background, and
$\delta(t)$ for combinatorial background.
Parameter $\tau^{}_{D3b}$ is obtained from fitting sideband 
data, and $\tau^{}_{Dch}$ is obtained from MC simulation.
These time distributions are convolved 
with resolution functions. For random-$\pi$ background 
(and the small multi-body $D^0/D^+/D^+_s$ background),
the resolution function used is the same as that for signal; 
for combinatorial background, the resolution function 
has the same form but parameters 
$f^{\rm (comb)}_{\rm tail}\!,\,S^{\rm (comb)}\!,\,S^{\rm (comb)}_{\rm tail}$
are determined from fitting sideband data.

The fitting procedure is implemented in steps as follows.
We first fit the RS signal region using a simple background
model to obtain a first estimate of signal resolution 
function parameters. We use this resolution function 
to fit an RS sideband region, which yields
$f^{\rm (comb)}_{\rm tail}$, $S^{\rm (comb)}$, $S^{\rm (comb)}_{\rm tail}$,
and $\tau^{}_{D3b}$ for RS background. 
We then fit the RS signal region with these parameters fixed, 
which yields $\mu$, $f^{}_{\rm tail}$, $S$, $S^{}_{\rm tail}$
and, as a check,~$\tau^{}_{D^0}$. We subsequently use this 
resolution function to fit the WS sideband region, obtaining 
$f^{\rm (comb)}_{\rm tail}$, $S^{\rm (comb)}$, $S^{\rm (comb)}_{\rm tail}$,
and $\tau^{}_{D3b}$ for WS background. Finally, we fit the WS signal 
region, fixing these background parameters and those of the signal 
resolution function; this yields $R^{}_D,\,x'^2$, and~$y'$.

The above fitting procedure has undergone several checks.
In MC simulation, background parameters obtained from the sideband region 
fit describe well the background in the signal region.
The resolution function obtained from data is very similar to
that obtained from the~MC.
The lifetime $\tau^{}_{D^0}$ obtained from the RS signal region
fit is $415.1\pm 1.4$~fs, consistent with the PDG value~\cite{pdg};
the $\chi^2$ of the fit projection is 56.0 for 55 bins.
We  generated MC samples having the same size as 
the data sample, adding the corresponding amount of background, 
and repeated the fitting procedure. For nine sets of
$(x'^2,\,y')$ values (spanning the ranges 
$[\,0,2\,]\times 10^{-3}$ and $[-2,2\,]\times 10^{-2}$, respectively), 
the fit recovers values consistent with the generated values. For 
these sets we also generated ensembles of ``toy'' MC
experiments, i.e., without detector simulation,
and smeared the time distributions appropriately. 
Fitting these experiments shows negligible fit bias.

\begin{figure}
\vskip0.30in
\centerline{\epsfxsize=10.0cm \epsfbox{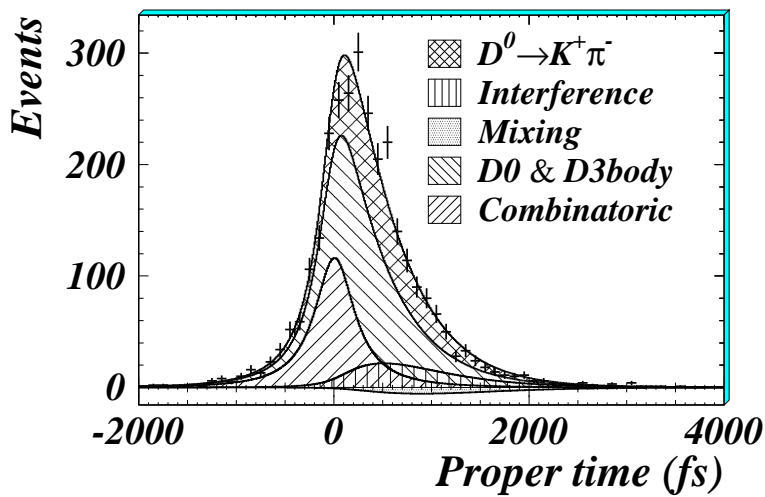}}
\vskip-0.10in
\caption{The decay-time distribution for WS events
satisfying $\left|m^{}_{K\pi}-m^{}_{D^0}\right| < 22$\mevm\ 
and $\left|\,Q - 5.9{\rm\ MeV}\right| < 1.5$\meve.
Superimposed on the data (points with error bars) 
are projections of the decay-time fit.}
\label{fig:time_fit}
\end{figure}

To this point in the analysis, all optimization of selection 
criteria was done ``blindly,'' i.e., without fitting WS signal 
events. We now fix the criteria and fit this sample. Four 
fits are done, yielding the results 
listed in Table~\ref{tab:fit_results}.
For the first fit we require that \cp\ be
conserved. The projection of this fit 
superimposed on the data is shown in 
Fig.~\ref{fig:time_fit}; the $\chi^2$ of
the projection is 71.9 for 60 bins.
The central value for $x'^2$ 
is negative (i.e., outside the physical region); 
thus the most-likely value is zero, and 
we refit the data fixing $x'^2=0$. 
The $\chi^2$ of this fit projection 
is 73.2 for 60 bins, which is satisfactory.
The $y'$ value is $\sim$\,$2\sigma$ from 
zero; when we generate
MC experiments with this value
(and $x'^2\!=\!0$), we find that the probability 
of obtaining an $x'^2$ value as negative 
as what we measure in the data is~8\%.
For the third fit we allow for \cpv\ 
and fit the \dkpiw\ and $\dbar\ra K^-\pi^+$ samples separately.
For $R^+_D$ and $R^-_D$ we obtain
$(0.255\,^{+0.058}_{-0.056})$\% and 
$(0.324\,\pm\,0.052)$\%, respectively.
We calculate $A^{}_D$ and $A^{}_M$ and 
use Eqs.~(\ref{eqn:cpv1}) and 
(\ref{eqn:cpv2}) to solve for $x'^2$ and $y'$.
Finally, for the last fit we assume no mixing or \cpv\ 
and set $x'^2\!=\!y'\!=\!0$;
the $\chi^2$ of this fit projection is 75.6 for 
60 bins, somewhat worse than for the case of mixing.

\begin{table}[hbt]
\caption{Summary of results from the separate likelihood fits.
The 95\% CL intervals are obtained from a frequentist method 
(see text) and include systematic errors.}
\centering
\begin{tabular}{c|ccc}
\hline
Fit Case & Parameter & Fit Result & 95\% CL interval\\
       &   &   $(\times 10^{-3})$ & ($\times 10^{-3})$ \\
\hline
         &  $x'^2$ & $-1.53\,^{+0.80}_{-1.00}$  &  $x'^2\!<\!0.81$ \\ 
No $CPV$ &  $y'$ & $25.4\,^{+11.1}_{-10.2}$     &  $-8.2\!<\!y'\!<\!16$ \\
         & $R^{}_D$ & $2.87\pm 0.37$            &  $2.7\!<\!R^{}_D\!<\!4.0$ \\
         &  $R^{}_M$ & -- & $R^{}_M < 0.42$ \\
\hline
No $CPV$ &  $y'$ & $6.0\pm 3.3$ & -- \\
$x'\!=\!0$ (fixed)   & $R^{}_D$ & $3.43\pm 0.26$ & -- \\
\hline
   & $A^{}_D$ & $-80\,\pm\,77$ & $-250\!<\!A^{}_D\!<\!110$ \\
$CPV$ allowed & $A^{}_M$ & $987\,^{+13}_{-380}$ & $-991\!<\!A^{}_M\!<\!1000$ \\
   &  $x'^2$ & -- & $x'^2\!<\!0.89$  \\ 
   &  $y'$   & -- & $-30\!<\!y'\!<\!27$ \\
   &  $R^{}_M$ & -- & $R^{}_M < 0.46$ \\
\hline
\begin{tabular}{c} No mixing \\ or $CPV$ \end{tabular}
& $R^{}_D$ & \multicolumn{2}{c}{
$3.81\,\pm\,0.17\,({\rm stat.})\,^{+0.08}_{-0.16}\,({\rm syst.})$} \\
\hline
\end{tabular}
\label{tab:fit_results}
\end{table}

\begin{figure}
\centerline{\epsfxsize=11.0cm \epsfbox{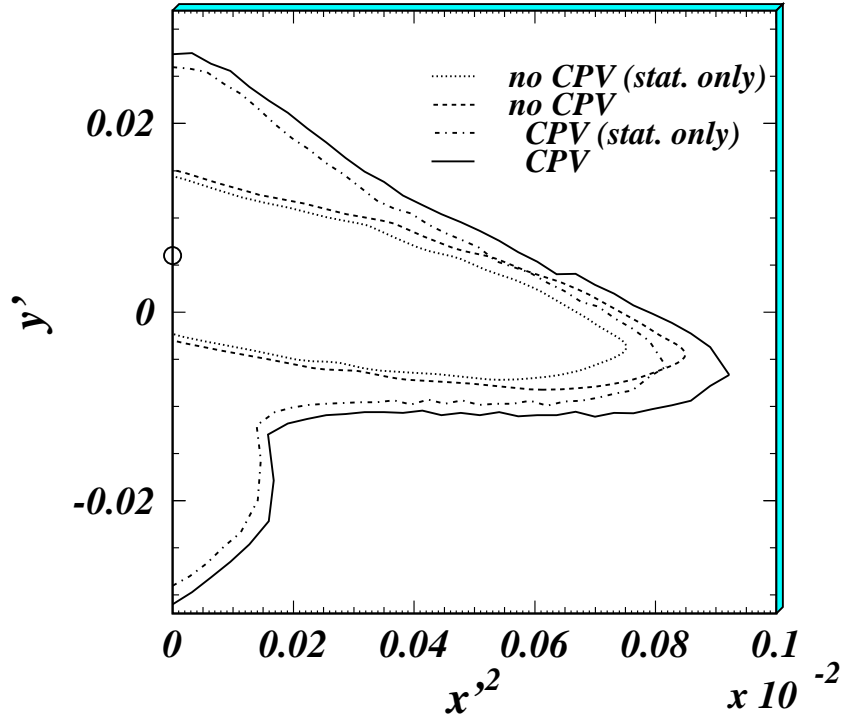}}
\caption{95\% CL regions for $(x'^2,\,y')$. 
The dotted (dashed) contour is statistical 
(statistical and systematic) and corresponds to 
\cp\ conservation. The dash-dotted (solid) contour 
is statistical (statistical and systematic) and 
allows for \cpv. The open circle represents the 
most-likely value when \cp\ is conserved and 
$x'^2$ is constrained to be $\geq$\,0.}
\label{fig:contours}
\end{figure}

To obtain 95\% CL limits on $x'^2$ and $y'$, we 
use a frequentist method (similar to that used in 
Ref.~\cite{babar}) with Feldman-Cousins 
ordering~\cite{FeldCous}. For points 
$\vec{\alpha}=(x'^2,\,y')$, we generate ensembles
of toy MC experiments and fit them using
the same procedure as that used for the data. For each
experiment we record the difference in likelihood 
$\Delta L = \ln L^{}_{\rm max} - \ln L(\vec{\alpha})$,
where $L^{}_{\rm max}$ is evaluated for $x'^2\geq 0$.
The locus of points $\vec{\alpha}$ for which 95\% of 
the ensemble has $\Delta L$ less than that of the data 
is taken as the 95\% CL contour.
This contour is shown in Fig.~\ref{fig:contours}.

To allow for \cpv, we obtain separate
$1\!-\!\sqrt{0.05}\!=\!77.6\%$ CL contours for 
$(x'^{+\,2},\,y'^+)$ and $(x'^{-\,2},\,y'^-)$. We 
combine points on the $(x'^{+\,2},\,y'^+)$ contour with 
points on the $(x'^{-\,2},\,y'^-)$ contour and use these 
values to solve Eqs.~(\ref{eqn:cpv1}) and (\ref{eqn:cpv2})
for $x'^2$ and $y'$. Because the relative sign of
$x'^+$ and $x'^-$ is unknown, there are two
solutions (one for each sign); we plot both
in the $(x'^2,\,y')$ plane and take
the outermost envelope of points to be the 95\% CL
contour allowing for~\cpv. This contour has a 
complicated shape due to the two solutions.
Because the contour includes the point 
$x'^2\!=\!y'\!=\!0$, we cannot constrain 
$\phi$ at this C.L.

We evaluate systematic errors by varying
parameters used to select and fit the data
within their uncertainties and recording the new  
fit values $\vec{\alpha}^{}_{\rm new}$ obtained.
These values are shifted with respect 
to the original central value $\vec{\alpha}^{}_0$.
We find the significance $m$\/ of a shift via the formula
$m^2\!=\!-2[\,\ln L(\vec{\alpha}^{}_{\rm new})\!-\!\ln L
(\vec{\alpha}^{}_0)]/2.30$,
where the factor 2.30 corresponds to 
68.3\% confidence in two dimensions.
We add in quadrature the significances of 
individual shifts to obtain an overall scaling factor
$\sqrt{1+\sum m^2_i}$. We increase the 95\% CL 
statistical contour by this factor to include
systematic errors. As a check, we generate an ensemble
of toy MC experiments with $\vec{\alpha}^{}_{\rm MC}\!=\!(0.,0.006)$
and fit them to confirm that
68.3\% of the ensemble satisfies 
$-2[\,\ln L(\vec{\xi})\!-\!\ln L(\vec{\alpha}^{}_{\rm MC})]\!<\!2.3$,
where $\vec{\xi}$ is the central value for an experiment.

The parameters varied include kaon and pion
identification criteria, the 
allowed
$\chi^2$ of the vertex fit,
background shape and normalization parameters, and resolution 
function parameters for both signal and combinatorial background.
The largest shift in $(x'^2,\,y')$ occurs for the
$D^{*\,+}$ momentum cut; when this is varied over
a significant range, 
$(\Delta x'^2/x'^2)^{}_{\rm max} = 12\%$,
$(\Delta y'/y')^{}_{\rm max} = 10\%$, and
$\Delta (-2\ln L) = 0.092$. The overall
scaling factor is $\sqrt{1+\sum m^2_i }=1.08$.
For the general case allowing for \cpv, we
scale the $D^0$ and $\dbar$ contours separately
before combining. The rescaled 95\% CL contours 
are shown in Fig.~\ref{fig:contours}:
the dashed contour corresponds to the \cp-conserving case 
and the solid contour to the general case.
Projecting these contours onto the coordinate axes 
gives the 95\% CL intervals for $x'^2$ and $y'$ 
listed in Table~\ref{tab:fit_results}. 

In summary, we have searched for $D^0$-$\dbar$ mixing and 
\cp\ violation using WS \dkpiw\ decays. In 90~fb$^{-1}$ 
of data we find no evidence for these processes and constrain 
the mixing parameters $x'^2$ and $y'$ and the \cp\ asymmetry 
parameters $A^{}_D$ and~$A^{}_M$. The limits for $x'^2$ and $y'$ 
are more stringent than previous results; the value for $R^{}_D$ 
(the ratio of DCS to CF decays) has smaller uncertainty.

We thank the KEKB group for the excellent
operation of the accelerator, the KEK cryogenics
group for the efficient operation of the solenoid,
and the KEK computer group and the NII for valuable computing and
Super-SINET network support.  We acknowledge support 
from MEXT and JSPS (Japan); ARC and DEST (Australia); NSFC (contract
No.~10175071, China); DST (India); the BK21 program of MOEHRD and the
CHEP SRC program of KOSEF (Korea); KBN (contract No.~2P03B 01324,
Poland); MIST (Russia); MESS (Slovenia); NSC and MOE (Taiwan); 
and DOE (USA).


\begin{thebibliography}{99}
\bibitem{cicerone} For a comprehensive review see S.\ Bianco, F.\,L.\,Fabbri, 
D.\,Benson, and I.\,Bigi, Riv.~Nuovo Cim.~{\bf 26N7-8}, 1 (2003).
See also A.~Falk {\it et al.}, Phys.\,Rev.~D {\bf 69}, 114021 (2004). 
\bibitem{kekb}
	S.~Kurokawa and E.~Kikutani, Nucl.\,Instr.\,Meth.~A {\bf 499}, 1 (2003).
\bibitem{belle_detector} A.\,Abashian {\it et al.\/} (Belle), 
			Nucl.\,Instr.\,Meth.~A {\bf 479}, 117 (2002). 
\bibitem{charge-conjugate}
		Charge-conjugate modes are included throughout this paper 
		unless noted otherwise.
\bibitem{e791} E.\ Aitala {\it et al.\/} (E791),
		Phys.\,Rev.~D {\bf 57}, 13 (1998). 
 \bibitem{cleo} R.\,Godang {\it et al.\/}\,(CLEO),\,Phys.\,Rev.\,Lett.\,{\bf 84},\,5038\,(2000).
\bibitem{babar} B.\,Aubert {\it et al.\/}\,(BaBar),\,Phys.\,Rev.\,Lett.\,{\bf 91},\,171801\,(2003).
\bibitem{pdg} S. Eidelman {\it et al.\/} (PDG), Phys.\ Lett.~B {\bf 592}, 1 (2004).
\bibitem{FeldCous} G.\,J.\,Feldman and R.\,D.\,Cousins,\,Phys.\,Rev.~D\,{\bf 57},\,3873\,(1998). 
\end{thebibliography}
\end{document}